\documentclass[sigchi,nonacm]{acmart}

\AtBeginDocument{%
  \providecommand\BibTeX{{%
    \normalfont B\kern-0.5em{\scshape i\kern-0.25em b}\kern-0.8em\TeX}}}

\usepackage{subcaption}
\usepackage{multirow}
\usepackage[inline]{enumitem}

\definecolor{darkgreen}{RGB}{24, 130, 0}

\setcopyright{none}
\begin{document}

\title{T-RECS: A Simulation Tool to Study the Societal Impact of Recommender Systems}

\author{Eli Lucherini}
\affiliation{%
  \institution{Princeton University}
  \city{Princeton}
  \state{NJ}
  \country{USA}}
\email{elucherini@cs.princeton.edu}

\author{Matthew Sun}
\affiliation{%
  \institution{Princeton University}
  \city{Princeton}
  \state{NJ}
  \country{USA}}
\email{mdsun@princeton.edu}

\author{Amy Winecoff}
\affiliation{%
  \institution{Princeton University}
  \city{Princeton}
  \state{NJ}
  \country{USA}}
\email{aw0934@princeton.edu}

\author{Arvind Narayanan}
\affiliation{%
  \institution{Princeton University}
  \city{Princeton}
  \state{NJ}
  \country{USA}}
\email{arvindn@cs.princeton.edu}

\renewcommand{\shortauthors}{Lucherini, Sun, Winecoff, and Narayanan}

\begin{abstract}
  Simulation has emerged as a popular method to study the long-term societal consequences of recommender systems. This approach allows researchers to specify their theoretical model explicitly and observe the evolution of system-level outcomes over time. However, performing simulation-based studies often requires researchers to build their own simulation environments from the ground up, which creates a high barrier to entry, introduces room for implementation error, and makes it difficult to disentangle whether observed outcomes are due to the model or the implementation.

  We introduce T-RECS\footnote{\href{https://github.com/elucherini/t-recs}{https://github.com/elucherini/t-recs}}, an open-sourced Python package designed for researchers to simulate recommendation systems and other types of sociotechnical systems in which an algorithm mediates the interactions between multiple stakeholders, such as users and content creators. To demonstrate the flexibility of T-RECS, we perform a replication of two prior simulation-based research on sociotechnical systems. We additionally show how T-RECS can be used to generate novel insights with minimal overhead. Our tool promotes reproducibility in this area of research, provides a unified language for simulating sociotechnical systems, and removes the friction of implementing simulations from scratch.

\end{abstract}
\maketitle
\pagestyle{plain}

\section{Introduction} \label{sec:intro}
Recommender systems in social media platforms such as Facebook and Twitter have been criticized due to the risks they might pose to society. For example,  ``filter bubbles" \cite{pariser2011filter} have been associated with the emergence of echo chambers leading to degraded political discourse online. Similarly, there is evidence that false news might spread faster and wider online than the truth \cite{vosoughi2018spread}, with the phenomenon potentially exacerbated by recommendation algorithms \cite{holmes2016}. YouTube has come under fire for its potential links to the radicalization of certain audiences \cite{tufekci2018} such as young voters in Brazil \cite{fischer2019}. However, there is no consensus on the scope of these concerns or ways to remedy them. For example, several studies have argued that the contribution of algorithmic systems to echo chambers is minimal or nonexistent \cite{hannak2013measuring,bakshy2015exposure,fletcher2018automated}.

Because phenomena such as filter bubbles arise through repeated system interactions over time, methods that assess the system at a single time point provide minimal insight into the mechanisms behind them. In contrast, simulations can model how system elements such as user, items, and algorithms interact over arbitrarily long timescales. As a result, simulation has proved to be a valuable tool in assessing the impact of recommendation systems on the content users consume and on society \cite{yao2017beyond, chaney2018, ciampaglia2018algorithmic, perra2019modelling, geschke2019triple, aridor2020deconstructing}. Most simulation studies of algorithmic systems have relied upon ad-hoc systems for implementation, which presents several challenges for the advancement of scientific understanding of the effects of algorithmic system dynamics.

We present T-RECS (Tools for RECommender system Simulation), an open-source, unified simulation tool designed to enable investigations of emerging complex phenomena caused by millions of individual actions and interactions in algorithmic systems including filter bubbles, political polarization, and (mis)information diffusion. 

T-RECS can mitigate problems associated with ad-hoc systems in several ways. First, one of the most time and labor intensive components of developing simulations is the engineering effort necessary to build the system. T-RECS has been developed to drastically reduce the engineering effort needed to develop a recommender system simulator. This allows researchers to shift their focus from the mechanics of the simulations to the scientific assumptions behind them. As a result, T-RECS can accelerate the pace of development and can facilitate high-quality simulation studies. Second, because many system elements in T-RECS have been designed with appropriate checks, its use can reduce the likelihood of software bugs that either slow the research process or contribute to erroneous results. Third, T-RECS provides a simple modular programming interface and terminology intuitive to researchers in the social sciences and computer science.

T-RECS is an easy-to-use but powerful framework; using the pre-populated library of common recommender system models and other system elements, researchers can instantiate a simulation using three lines of code. For instance, the following code runs a simulation with 1,000 users and 10,000 items whose interactions are mediated by the system default content-filtering algorithm (see Section \ref{sec:design}). The last line gathers default measurements.

\begin{verbatim}
recsys = trecs.models.ContentFiltering(num_users=1000, 
                                       num_items=10000)
recsys.run(timesteps=100)
measurements = recsys.get_measurements()
\end{verbatim}

With T-RECS, researchers with expertise in social science but limited engineering expertise can still leverage simulation to answer important questions about the societal effects of algorithmic systems. Yet, the framework also provides flexibility for expert users to build upon. For example, researchers can easily specify each user and item's representation in the system, along with new custom measurements.

Applying the same tool to different problems 
\begin{enumerate*}[label={(\arabic*)}]
    \item promotes reproducibility by allowing researchers to easily share their simulations;
    \item provides a common language to describe problems in the literature; and
    \item fosters discovery of principles that apply across seemingly different problems.
\end{enumerate*}

In addition to social science and algorithmic researchers, T-RECS can also inform the work of engineers, who can compare algorithmic design choices, as well as policymakers, who can use the results to regulate potentially harmful phenomena.
\section{Background and related work} \label{sec:background}

T-RECS is a tool for evaluating recommendation systems and other algorithmic information systems, especially their societal consequences, based on agent-based modeling. We briefly review the relevant background.

\subsection{Recommender systems}
Recommender systems are responsible for a large part of the content we see and interact with online. For example, in 2018 YouTube reported that 70\% of the views were derived from recommendations \cite{solsman2018}. In 2019, over half of the more than 1 billion daily active users on Instagram turned to Explore, the recommendation-driven section of Instagram, at least once a month \cite{medvedev2019}; and numbers from 2016 suggest that 80\% of the hours streamed by users on Netflix were driven by its recommender system \cite{gomezuribe2016}. Given how frequently recommendations influence our day-to-day lives, understanding how recommendation algorithms influence both short-term and long-term outcomes for users is critical.

The algorithms undergirding recommender systems can take advantage of a variety of information to serve recommendations to users. Collaborative filtering is a subset of recommendation algorithms that leverages patterns in existing user interaction data to generate predictions for new items. Methods such as user-based or item-based collaborative filtering use a nearest-neighborhood approach whereas matrix factorization collaborative filtering uses latent representations of users and items to make predictions. In contrast to wholly interaction-based methods, content-based algorithms take advantage of item or user meta-data (e.g., genre, director for movies). More recently, Bayesian approaches (e.g., \cite{rendle2012bpr}), reinforcement learning (e.g., \cite{zheng2018drn}), and deep learning (e.g., \cite{wang2015collaborative, cheng2016wide, zheng2017joint}) that draw from methods in machine learning more generally have also been applied to recommendation problems. Hybrid and ensemble approaches leveraging multiple recommendation models are often used to mitigate weaknesses associated with a single algorithm type \cite{aggarwal2016ensemble}.

Recommender systems are typically optimized for predictive accuracy \cite{herlocker2004evaluating}, but often include additional metrics such as diversity and novelty of content \cite{ziegler2005improving, hurley2011novelty} and, more recently, fairness-related metrics \cite{burke2018balanced, kamishima2018recommendation, dean2020recommendations}.

\subsection{Recommender systems and society}

Recommender systems help users find content that better matches their short or long-term preferences, and to find it more efficiently. They also help users discover and develop new interests. From the platform's perspective, recommender systems can influence user behavior in service of the platform's goals (such as increasing user engagement), improve the visibility of new products, and provide a mechanism for learning more about users' preferences \cite{jannach2016recommendations}.

On the other hand, there are a number of potentially harmful effects. \textbf{Filter bubbles} are the result of recommender systems presenting a user with content similar to the user's history, resulting in intellectual isolation, political \textbf{polarization}, and \textbf{echo chambers} \cite{sunstein_2001, sunstein_2009, sunstein2009going, pariser2011filter}. A more extreme version is \textbf{radicalization}, a feedback loop in which the algorithm encourages users to consume more and more extreme content, gradually nudging them to embrace increasingly radical ideas  \cite{munn2019alt}. 

Recommender systems may exhibit a bias toward popular items, resulting in \textbf{homogenization} of user behavior and a \textbf{concentration} of the market for content in the hands of a small number of creators \cite{chaney2018, ciampaglia2018algorithmic, salganik2006experimental}. 

Finally, recommender systems may contribute to the spread of misinformation online. These phenomena are inter-related: for example, misinformation may be more likely to spread in the presence of echo chambers because users are less likely to encounter corrective information.

Although we have characterized these potential harms as effects of recommender systems, they may also raise from other types of online information systems including search \cite{golebiewski2018data} and social networks \cite{bikhchandani1998learning}. In the shift from offline to online information seeking, users are freed from physical-world constraints such as word-of-mouth recommendation networks or the limited content available through a newspaper subscription. Online, users can much more easily find content tailored to their preferences, beliefs, and ideology. 

Our understanding of the long-term impact of recommender systems on users and society is still nascent, and there is little consensus in the literature. For example, quantitative research on the effects of algorithmic systems has found little evidence of filter bubbles \cite{hannak2013measuring,bakshy2015exposure,fletcher2018automated}, in contrast to other types of inquiry such as ethnography \cite{golebiewski2018data, tripodi2018searching, lewis2018alternative}. Even studies with the same broad methodology produce conflicting findings, such as empirical studies of misinformation \cite{vosoughi2018spread,friggeri2014rumor} and simulation studies of echo chambers \cite{aridor2020deconstructing, geschke2019triple}.

Many potential factors could explain these differences. Echo chambers or other effects may exist but not be caused by the recommender system, or the effects may only become apparent over time due to feedback loops and not be revealed by cross-sectional studies, or there may be differences between platforms. A further wrinkle related to platform differences is that some studies cannot be reproduced externally because the data is proprietary.

\subsection{Agent-based modeling}

Agent-based models help researchers understand the properties of a complex system by modeling of the individual agents that comprise the system and the interactions between them \cite{bonabeau2002agent}. One of the earliest Agent-Based Models (ABMs) in the social sciences is Thomas Schelling's work on the dynamics of racial segregation about fifty years ago \cite{schelling1971dynamic}. ABMs became widespread only after the rise in the availability computational power  of the 1990s \cite{epstein1996growing, lorek1999modelling, axelrod1997complexity}. ABMs are widely used in areas of research such as economics and finance, ecology, biology, and epidemiology \cite{heard2015agent}. 

\subsection{Content creators}
Academic research on recommender systems have often focused on user-centric experiences and outcomes \cite{abdollahpouri2020multistakeholder}. However, recommender systems often serve multiple stakeholders including platform providers and content creators. Increasingly, researchers have recognized the importance of a multistakeholder perspective for developing and evaluating recommender systems \cite{abdollahpouri2020multistakeholder, burke2017multisided, munger2020right}. Indeed, content creators themselves have raised questions about how recommendation algorithms affect their livelihoods or entrench societal biases \cite{mccluskey.2020}. Moreover, content creators represent a dynamic set of agents, and they may shift and adapt in response to algorithmically-mediated interactions with users \cite{cotter2019playing, ben2018game}. 

So far, most simulation-based research on the system-level effects of algorithmic system features have also focused on users and have likewise failed to account for either the dynamic effects of content creators or the consequences of system properties for content creators. 




\section{The Case for T-RECS} \label{sec:methodology}
\begin{table*}[t]
\begin{tabular}{lllll}
\toprule
\multicolumn{5}{c}{\textbf{Simulation to study sociotechnical systems}} \\
\multicolumn{1}{l}{Work}                          & \multicolumn{1}{l}{Model} & \multicolumn{1}{l}{Main topic} & \multicolumn{1}{l}{Tool} & \multicolumn{1}{l}{ABM?}      \\ \hline
Aridor et al.~\cite{aridor2020deconstructing}                  & RS & Filter bubbles & n.a. & Yes        \\
Chaney et al.~\cite{chaney2018}                  & RS & Filter bubbles & n.a. & Yes     \\
Ciampaglia et al.~\cite{ciampaglia2018algorithmic}              & RS  & Popularity bias & n.a.  &  No     \\
Garimella et al.~\cite{garimella2017balancing}               & ID & Filter bubbles &  n.a. & Yes \\
Geschke et al.~\cite{geschke2019triple}                 & RS & Filter bubbles &  NetLogo \cite{netlogo} &  Yes      \\
Goel et al.~\cite{goel2016structural}                    & ID & Virality & n.a. &  Yes \\
Jiang et al.~\cite{jiang2019degenerate}                   & RS & Filter bubbles &  n.a. & No  \\
Lee et al.~\cite{lee2019homophily}                     & OD & Perception biases & n.a. & Yes \\
Lim et al.~\cite{lim2014opinion}                     & ID & Digital divide & NetLogo & Yes \\
Nasrinpour et al. \cite{nasrinpour2016agent}                 & ID & Virality &  AnyLogic \cite{borshchev2014multi}  & Yes \\    
Perra and Rocha \cite{perra2019modelling}                   & OD  & Polarization &  n.a. & No  \\
Sun et al.~\cite{sun2019debiasing}                     & RS &  Popularity bias &  n.a.  & Yes      \\
Tambuscio et al.~\cite{tambuscio2018network}               & ID  & Misinformation &  n.a. & Yes     \\
T\"ornberg \cite{tornberg2018echo}     & ID & Misinformation & n.a. & Yes \\
Yao and Huang \cite{yao2017beyond}                 & RS & Popularity bias &  n.a.  & Yes \\  
\bottomrule
\end{tabular}
\caption{Simulation literature survey. RS = recommender systems, ID = online information diffusion, OD = online opinion dynamics}
\label{tab:survey}
\end{table*}
As noted in Section \ref{sec:background}, there are still many open questions on the long-term impact of recommender systems on users, content providers, and society at large. T-RECS offers a unified tool to contextualize and understand seemingly contradictory results, such as those from the literature on filter bubbles. The design of T-RECS facilitates the implementation of related problems using the same simulation framework.

Similarly, T-RECS allows studying multistakeholder problems, such as the complex interactions between users and content creators when they are mediated by an algorithm that tends to suggest more and more radicalized content.

In this section, we explain what simulation, in particular agent-based modeling, can offer. Finally, we argue in favor of unified tools such as T-RECS.

\subsection{The case for simulation}

%

Observational and experimental methods have been widely used to study the impact of recommender systems \cite{nguyen2014exploring, ali2019discrimination, hosseinmardi2020evaluating, cheng2016wide}.
Contrary to observational methods, simulation provides the tools for generalizable and causal discoveries; furthermore, models are typically not affected by ethical issues that arise when experimenting on real users. Furthermore, while methods like offline evaluation on historical data (e.g., the MovieLens dataset \cite{harper2015movielens}) \cite{burke2018balanced, kamishima2018recommendation, dean2020recommendations}  are useful to evaluate algorithms on fixed or unknown user strategies, simulation provides an additional degree of control on evaluating systems on varying user behaviors. Finally, contrary to analytical approaches, simulation supports complex environments and focuses on the dynamics of the system instead of static points of equilibrium.



\subsection{The case for Agent-Based Modeling}

%

Agent-based modeling is useful to study emergent phenomena: those that arise from the interactions of individual agents, often in counter-intuitive ways. Bonabeau gives the example of a traffic jam, noting that it may move in the opposite direction from the cars that are causing it \cite{bonabeau2002agent}. He presents a list of system properties that may give rise to emergent phenomena, such as adaptive agents and network effects, all of which are present in recommender systems.

Recently, ABMs have been explicitly used to describe interactions mediated by a recommender system. For example, Geschke et al.~\cite{geschke2019triple} modeled an online information diffusion network with users and pieces of information as the agents; similarly, Nasrinpour et al.~\cite{nasrinpour2016agent} studied message propagation on Facebook with an ABM. For more examples, see Table \ref{tab:survey}.


ABMs are especially apt at discovering indirect and unintended effects of design choices because the outcomes of simulations emerge from interactions. Therefore, researchers do not need to encode their expectations into the initial assumptions \cite{de2014agent}.

Agent-based modeling can express geographical and social distance \cite{de2014agent}. Thus, they are suitable to model \textit{adjacency networks} between users, which are often associated with recommender systems. For example, recommender systems with a community-based component leverage connections between users and their friends to provide recommendations \cite{ricci2011introduction}.
%

\subsection{The case for a new tool}
To determine whether a new tool can be useful, we surveyed existing simulation studies of algorithmic systems, finding 15 studies. A summary of the studies we reviewed is in Table \ref{tab:survey}.


We argue in favor of unified tools in this space, such as T-RECS, with the following observations. 
First, these 15 studies represent recently published usages of simulation to understand societal impact of algorithmic systems. Of these, seven specifically focus on generating knowledge on recommender systems, six focus on online information diffusion, and two focus on opinion dynamics. 

However, all of the studies we reviewed involved system components conceptually representing ``users,'' ``items,'' and ``algorithms.'' As a result, we designed T-RECS to also have system elements that could represent these types of entities and enable flexible representations of their properties and behaviors. 

Second, our literature review indicates that simulations of algorithmic systems often rely on ad-hoc software, which can be challenging to develop and reproduce. For example, 12 out of the 15 studies in Table \ref{tab:survey} were developed independently, including six of the seven studies focusing on recommender systems. Two used NetLogo, a programming language and modeling environment originally developed in 1999 for multi-agent simulations, and one used AnyLogic, which is proprietary GUI-based simulation software available for academic use \cite{tisue2004netlogo, borshchev2014multi}. The heterogeneity in implementations in the studies points to the need for a unified framework.

The primary benefits of applying the same tool to different problems are both practical and scientific. From the practical perspective, a unified framework will reduce the engineering effort needed to develop a simulation, allowing researchers to shift focus from the mechanics of the simulation to the assumptions behind them. In addition to enabling researchers to focus more on the science of simulations, a common framework will also speed up development, facilitating a greater volume of high quality research on algorithmic systems. For example, if much of the up-front engineering effort has already been accomplished through the unified framework, there are fewer new system elements and consequently, fewer places where software bugs could lead to slow downs in development, or worse, erroneous results.

From the scientific perspective, a major issue with ad-hoc systems is that results can be difficult to reproduce. Because many idiosyncrasies in design and implementation are eliminated in a unified system, the likelihood that another research team could reproduce a scientific finding increases. Furthermore, a unified framework will allow researchers to communicate in a common language for different problems. For example, our literature review in Table \ref{tab:survey} demonstrates that multiple researchers have conducted simulations on filter bubbles; however, each of these studies has a unique definition for what constitutes a filter bubble and unique metrics for evaluating filter bubble effects. As a result, it is difficult to reconcile results across similar simulations, much less across simulations that address more disparate concepts. 

Before T-RECS, there have been other efforts to provide a unified simulation environment for sociotechnical systems \cite{tisue2004netlogo, bountouridis2019siren, d2020fairness, krauth2020offline, borshchev2014multi}. \textbf{NetLogo} was designed with a heavy emphasis on visualizing agents in 2D space and also requires users to learn the NetLogo programming language \cite{tisue2004netlogo}. Furthermore, NetLogo was not optimized for large-scale simulations for hundreds of thousands of users, and complex recommender system algorithms are not straightforward to implement in the NetLogo interface. \textbf{AnyLogic} is proprietary software for agent-based modeling developed by the AnyLogic company \cite{borshchev2014multi}. Simulations in AnyLogic are created through a graphical user interface, limiting expressiveness and flexibility. 

In the past few years, multiple simulation libraries specifically geared towards studying the temporal dynamics of simulated recommender systems have been released. \textbf{RecoGym}  was the first reinforcement learning simulation environment for recommendation, in the context of online advertising  \cite{rohde2018recogym}. The \textbf{RecSim} library was designed for a reinforcement learning-based approach to modeling recommender systems \cite{ie2019recsim}. Another follow-up simulation framework, \textbf{RecSim NG}, uses probabilistic programming for uncertainty modeling in recommender ecosystems \cite{mladenov2021recsim}. While these libraries offer useful mathematical tools for reasoning about recommender systems, they are geared towards expert recommender systems practitioners and researchers, requiring knowledge of libraries like Tensorflow, OpenAI Gym, or Edward2, and are designed for a particular technical frame, such as reinforcement learning or probabilistic programming. In contrast, T-RECS is designed to be flexible and simple enough that users familiar with Python and \texttt{numpy} will be able to build and run simulations relatively quickly, promoting accessibility to a broader audience of researchers, from computer scientists to social scientists. 

The most related tools to ours are \textbf{ML Fairness Gym} and \textbf{RecLab} \cite{d2020fairness, krauth2020offline}. \textbf{ML Fairness Gym} implements the OpenAI Gym API to provide a set of reusable components for studying long-run fairness in algorithmic decisionmaking systems over time. The ML Fairness Gym is framed to have a specific focus on fairness, while our tool is intended to allow practitioners to study a broad range of phenomena. Furthermore, the Fairness Gym does not have robust support for multi-agent simulation, instead focusing on the decision-maker as the primary agent. In contrast, T-RECS allows for dynamic behavior at the levels of users, recommender systems, and content providers. \textbf{RecLab}, a recently released Python library for simulating and evaluating recommender systems, is the most similar tool to T-RECS \cite{krauth2020offline}. While RecLab focuses on the evaluation of different recommender systems, T-RECS is designed to make it equally easy to study the effects of different user choice models, content creator behavior, or item distributions. As a result, T-RECS has the expressiveness to model sociotechnical systems beyond recommender systems, such as information diffusion in social networks.

\section{Design and architecture} \label{sec:design}
\begin{figure*}[!t]
\includegraphics[width=0.6\linewidth]{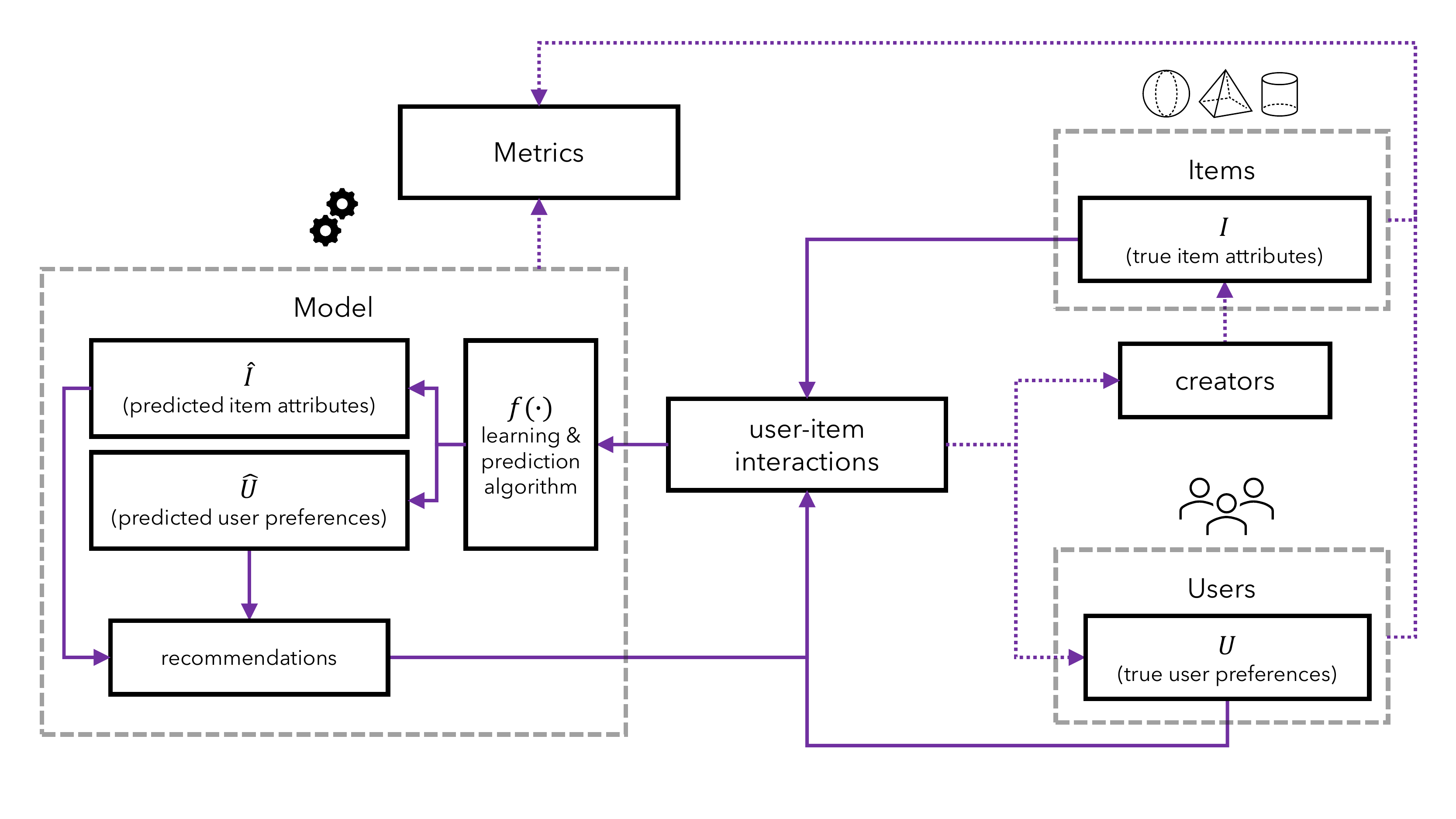}
\caption{A conceptual diagram indicating the different design components  of a T-RECS simulation, and the components that are used as inputs to other components as the simulation progresses. Lines with arrows indicate inputs fed into outputs. Dashed gray lines indicate conceptual components. Dotted arrows indicate possible feedback mechanisms (for example, item interactions affecting content creator distributions). }
\label{fig:arch_design}
\end{figure*}

Figure \ref{fig:arch_design} shows the modules of T-RECS, which include:
\begin{enumerate*}
    \item the \textbf{users}, representing the entities interacting with the recommendations: for example, consumers on an e-commerce platform;
    \item the \textbf{items}, representing the object of the recommendations: for example, the movies in a movie recommendation system; and the \textbf{content creators}, providing new items during the simulation;
    \item the \textbf{model} representing the mechanism through which users interact with items: for example, a collaborative filtering algorithm; and
    \item the \textbf{metrics} which evaluate the outcomes of the simulation sampled at each simulation step. Metrics also keep store information on selected \textit{internal states} of the system, such as the user preferences over time.
\end{enumerate*}
Note that these modules are common to many of the simulations in Table \ref{tab:survey}.
We go into more detail about each of the core modules in the subsequent sections.

\subsection{Simulation Dynamics} \label{sec:sec:dynamics}

We summarize the simulation dynamics of T-RECS, inspired by the model developed by Schmit and Riquelme \cite{schmit2018human}. We note that this simulation framework works for many algorithmic system that can be modeled as a \textit{user-algorithm feedback loop}. In the following sections, we present more details about how the basic dynamics can be enriched and changed. The numbers in the following list correspond to the numbers highlighted in Figure \ref{fig:arch_design}.

At each time step:
\begin{enumerate}
    \item The model predicts the user scores for each item in the system. These predicted scores, which can be different from the \textit{actual} user scores, are used by the system to make recommendations to users. By default, the predicted user scores are calculated as the inner product between user preferences and item attributes.
    \item The model presents a set of items to each user. In the general case, the system presents different items to each user, generally chosen considering the  user scores predicted in the previous step. At every step, items may also be randomly interleaved into recommendations, simulating user discovery of items outside of algorithmic recommendations, as in Chaney et al. \cite{chaney2018}. If runtime creation of items is enabled, the items most recently generated by the content creators will be included in system-generated recommendations based on how the specific recommendation algorithm handles new items. For example, in a popularity-based system, all new items have a popularity at zero and are placed near the bottom of recommendation lists.\footnote{T-RECS users may specify custom behavior for how new items are scored. For example, you may specify that all new items do not appear in recommendation lists and instead are randomly interleaved throughout the recommendations, as in Chaney et al. \cite{chaney2018}.}
    \item Each user gives feedback to the presented items. This means something different for different systems (e.g., in a movie recommender system, users choose movies to watch). Typical user feedback is \textit{implicit} (to simulate a click, generally only on one item). Users decide which item(s) to consume based on their actual preferences.
    \item The system records each user's feedback and updates its internal state. What this means varies with each model. For example, T-RECS' popularity model keeps track of the number of interactions each item has received; our content filtering model keeps track of the attributes each user has interacted with.
    \item The measurement module updates all tracked metrics based on the latest interactions. The default metrics vary by model, but could include an accuracy measure, such as the mean squared error between the predicted user scores and the actual user scores or a variety of other measures of recommendation list diversity, user interaction similarity, etc.
\end{enumerate}

\subsection{Users}
Users in T-RECS are represented by two main elements: their preferences and the score they attribute to each item in the system. With our design, informed by our survey of the literature, we assume that the real user preferences and the preferences predicted by the recommender algorithm are separate. Similarly, recommender algorithms may predict scores for the items in the system that do not perfectly correspond to the real scores attributed by the users. This modeling choice in several studies we analyzed in our survey \cite{chaney2018, aridor2020deconstructing}. Therefore, in T-RECS there are four dedicated data structures: \textit{actual} user preferences, \textit{predicted} user preferences, actual user scores, and predicted user scores.

Additionally, T-RECS provides a simple framework to change how user preferences evolve over time, typically in response to the items they are exposed to and consume. This behavior has been adopted in several of the studies we analyzed in our survey \cite{geschke2019triple, jiang2019degenerate}.

\subsection{Items and content creators}
Items in T-RECS can be fixed or dynamic. In a system such as a movie recommendation website, it might be useful to consider the catalog of items as fixed, since the changes to it are not necessarily dependent on the response of the audience to it.

However, in many real-world recommender systems the catalog of items is not fixed. Instead, it changes over time as producers, or \textit{content creators}, create and publish new items. For example, a platform such as YouTube might be better modeled with a dynamic catalog of items, as \textit{YouTubers} are as much a part of the ecosystem as viewers. Most importantly, creators are often incentivized to adapt the content they publish in direct response to the viewers' reactions. 
In T-RECS, we added a module for content creators to approximate this phenomenon. Researchers can define the incentives of the creators and how they change over time. 

Regardless of the specific item and content creator configuration, researchers can enforce additional constraints to how users consume items, or to how items are recommended to users. For example, it is possible for users to avoid consuming items they have already seen in the past; similarly, recommender models can avoid recommending items that the users have already interacted with.  

Content creators are characterized by the attributes of items they generate, and can be thought of as particular distributions from which items may be sampled. To mirror some of the complexity of content creator behavior in real systems, the attributes of the items content creators generate are sampled probabilistically.

\subsection{Model representation} \label{sec:sec:model}
%

In T-RECS, we make the following decisions regarding recommender system models:
\begin{enumerate}
    \item We use an \textit{event-driven} model as opposed to a \textit{time-driven} model--that is, we automatically skip all steps in which the system did not observe any user interactions. This increases the efficiency of the simulation. As a consequence, one simulation step does not correspond to a constant time unit.
    \item User interactions are processed in parallel. By default, at each step, the system receives one piece of feedback from each user, although T-RECS users can override this as necessary.
    Researchers who want to observe multiple user-item interactions before retraining a model can specify that behavior through the same framework. 
\end{enumerate}

Although our framework is particularly well-suited to simulating recommender systems, we find that it is also flexible enough to model a broader set of sociotechnical phenomena. For example, we used T-RECS to model the process of information diffusion in networks in which recommender systems do not necessarily have a role, as long as they follow the user-item-system feedback loop approach. In the context of information diffusion, users come into contact with various pieces of content (items) and probabilistically choose to share content with others in their network. At each step, the system monitors which users have shared content and presents shared content to their neighboring users at the next timestep, keeping track of which users have shared content (and thus will not reshare in the future). process of sharing  We offer more details on this topic in Section \ref{sec:sec:structural-virality}.

\subsection{Metrics} \label{sec:sec:metrics}
In T-RECS, metrics are designed for researchers to gather information about the model and the simulation. They often define the output of the simulation: for example, researchers interested in discovering filter bubbles can measure diversity of recommended content across users and draw conclusions from the results of these measurements.

Our main contribution in this regard is to provide a mechanism for easy implementation and  templates for researchers to use as an example. Our literature survey (Table \ref{tab:survey}) suggested very little consistency of metrics used across studies. While we could identify overarching themes --- the most prominent being distances between attributes and preferences --- each of the 14 studies we analyzed used distinct metrics. Therefore, we purposely provided a limited set of metrics with the expectation that researchers will want to implement their own metrics. 

\subsection{Programmer interface} \label{sec:sec:api}

Our programmer interface is beginner-friendly. We took inspiration from popular machine learning library SciKit-Learn \cite{scikit-learn} in that models can be instantiated and simulations can be run without specifying any argument. T-RECS was developed in Python and makes heavy use of the \texttt{numpy} library \cite{2020NumPy-Array}. We deliberately chose the default behaviors and provided an interface to easily modify them. For example, instantiating a simulation using the popularity model can be achieved by this line of code:

\begin{verbatim}
recsys = trecs.models.PopularityRecommender()
\end{verbatim}

The line above instantiates a popularity model with 100 users and 1250 items. User and item characteristics are also generated with default values from an arbitrary distribution (see the documentation for more details). We see these as sensible defaults, as the simulation is large enough to see emergent network effects while small enough to pose minimal computational burden; of course, T-RECS users are free to modify these parameters as they see fit.

Although T-RECS was designed to be accessible to beginners, it offers flexibility in changing a wide array of components for more advanced calibration. The customization is not limited to the parameters directly related to users and items. For example, researchers can define new score functions--that is,the functions calculating actual and predicted user scores. Instead of using the default inner product as the score function, the user can define their own mechanism for calculating user-item scores. As an example, we provide an alternative score function that calculates scores using the cosine similarity between $U$ (or $\hat{U}$) and $I$. We also provide a mechanism to define new score functions when initializing a model.

As with score functions, T-RECS offers an API to define new models, metrics, user feedback behaviors, and more.

\section{Case studies} \label{sec:case_studies}

Our goal in developing T-RECS was to create a tool that introduced a common conceptual framework for conducting simulations with different levels of complexity, different methodologies, and different goals, such as those enumerated in Table \ref{tab:survey}.  To illustrate that T-RECS has achieved this aim, we reproduced results from two substantially different simulation-based studies. In the first study, Chaney et al. \cite{chaney2018} examined how algorithmic confounding in recommendation systems leads to the homogenization of user behavior. In the second, Goel et al. \cite{goel2016structural} simulated models of contagion to understand the patterns of online information diffusion on Twitter. Although the two studies model distinct sociotechnical systems, both can be expressed within the T-RECS framework.

Our goal in reproducing these two studies was to perform a \textit{conceptual replication} rather than an \textit{exact replication} \cite{crandall_scientific_2016}. While exact replication aims to perform an identical operationalization of the original test as the original study, a conceptual study examines whether a hypothesis holds up under different operationalizations of the same conceptual variable. Conceptual replications provide evidence as to the robustness and generalizability of the theory. Although our replications are faithful to the original authors' descriptions of their methods, our goal was to show that T-RECS can be used to generate similar insights to those from these studies, rather than to generate wholly identical graphs or statistics. For this reason, we did not seek out source code from the authors of the original studies, but rather, reconstructed the simulation parameters within T-RECS using only the details provided in the published reports.

Both of the studies we replicated were originally implemented via ad-hoc systems. Without a common framework, a researcher interested in reproducing these studies would have to re-implement every element using the methodology presented in the papers, and ideally, would also have to implement all the necessary testing functionality from scratch. In contrast, because both replication studies involved representations for users, items, and models, we were able to use the core components of T-RECS as a starting point to reproduce the results of these prior studies rather than beginning anew. Similarly, we were able to take advantage of the many built-in testing capabilities of T-RECS for debugging. Relying on the core functionality of T-RECS allowed us to concentrate our efforts on faithfully reproducing aspects of our replication studies' designs rather than on engineering.

Our conceptual replication case studies demonstrate the power of T-RECS to accommodate complex and distinct algorithmic simulations; however, another major benefit of our framework is that simulation designs can be easily extended. In our third and final case study, we conduct a \textit{constructive replication}--a replication in which the findings of a prior study are replicated and then extended to encompass new elements that provide additional scientific insight. Our constructive replication study examines the homogenizing effect on item generation that is induced by the presence of content creators, who adapt over time to user feedback. For this study, we reuse much of the code and many of the assumptions from our replication of Chaney et al.'s research \cite{chaney2018}. As a result, we can confidently attribute any differences between the results of the simulations with and without content creators to the introduction of this new system element rather than faulty reproduction of the prior work's assumptions or implementation details.

In the following sections, we describe our motivations, methods, and results in adapting T-RECS to each case study. For more technical details, please see the Appendix. All code necessary to reproduce our experiments can be found online.\footnote{\href{https://github.com/sunnymatt/t-recs-experiments}{https://github.com/sunnymatt/t-recs-experiments}}

\subsection{Algorithmic confounding (Chaney et al.)}

\subsubsection{Background.} At a high level, Chaney et al. \cite{chaney2018} illustrate the detrimental effects of algorithmic confounding, which occurs when a recommendation algorithm is trained on user interaction data that is itself influenced by the prior recommendations of the algorithm. The study shows that algorithmic confounding homogenizes user behavior more than what would occur if all users were provided with recommendations that best matched their underlying preferences.

\subsubsection{Motivation.} We chose to replicate the results from Chaney et al. \cite{chaney2018} for several reasons. First, this investigation leveraged complex representations for each of the core components of the simulation (i.e., users, items, and models). Although the goal of the original study was not to approximate a real-world system at high fidelity, many of their choices of parameters and representation were based on an understanding of the characteristics of real-world systems (e.g., short head, long tail in item popularity). By illustrating that T-RECS can be modified to manifest this level of nuanced representation, we demonstrate the power of our simple architecture for encapsulating complicated specifications. That is, T-RECS can be used not only for simple toy models, but also for highly complex use cases. 
Second, this study focused on how the dynamics of one core component (models) could affect the outcome of the system over time through its influence on another component (users). Thus, by replicating this finding in T-RECS, we show that our system can be used to model interactions at the level of entities that ultimately affect the system as a whole. As this is generally the goal of most agent-based simulations, this suggests that our system will be well-suited to most questions of this nature. 

\subsubsection{Simulation overview} We begin with a high-level overview of the user-item-model interaction dynamics. Each user and each item has some ``true'' representation as a vector of attributes. At each timestep, every user is shown a list of items by the model, where the model attempts to show each user $u$ the set of items which have maximum predicted utility for $u$. Furthermore, at each timestep, ``new'' items are introduced into the system. These items are interleaved randomly into each user's recommendation list independently for each user. For each item $i$, there is a ``true utility'' that user $u$ obtains from interacting with that item that is a probabilistic function of $u$ and $i$'s attributes; however, each user has incomplete knowledge of their true utilities (i.e., they have an ``educated guess'' of how much they will like each item before interacting with it). Each user then interacts with one item on the basis of this imperfect knowledge of the utility that would be gained from each item in the recommendation list, plus an ``attention mechanism'' that accounts for the item's position in the recommendation list. The interaction data is then fed back into the model, which may or may not use the feedback to update its internal representation of users and items. After interacting with an item once, a user will not interact with that item again during the simulation.

Each model has its own method of internally representing users, items, and the predicted utility that each user $u$ obtains from interacting with each item $i$. For example, the popularity model represents each item $i$ as the number of times that any user has interacted with $i$, and the predicted utility for a given user $u$ and item $i$ is simply the popularity of item $i$ (i.e., all users are recommended the same set of the top-$k$ most popular items, setting aside new item interleaving). We implement and compare results between six different types of models: popularity, content filtering, matrix factorization, social filtering, random, and ideal.  The last two models are included for the purpose of comparison: the random model recommends items randomly to each user, while the ideal model recommends items on the basis of the true user-item utilities. For more details about each model in the simulation, see Section \ref{sec:appendix_replication}.

\begin{figure}[!t]
\begin{subfigure}{.5\textwidth}
    \includegraphics[width=\linewidth]{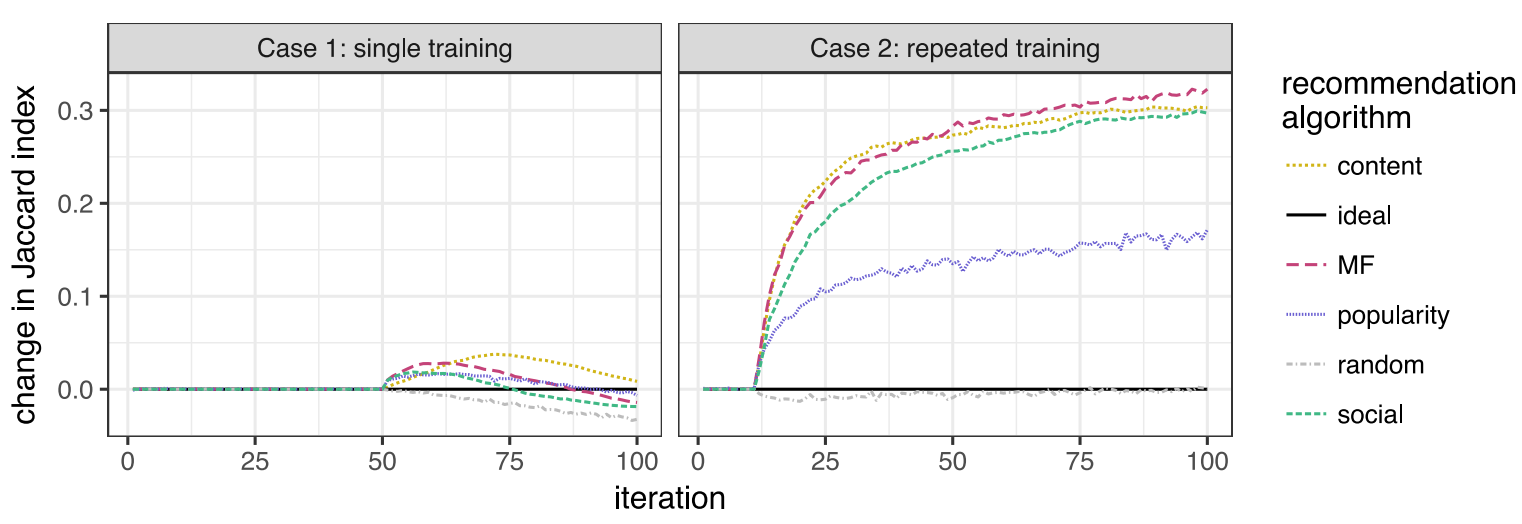}
    \caption{Original figure from Chaney et al \cite{chaney2018}}
    \label{fig:chaney_theirs}
\end{subfigure}
\begin{subfigure}{.5\textwidth}
    \includegraphics[width=\linewidth]{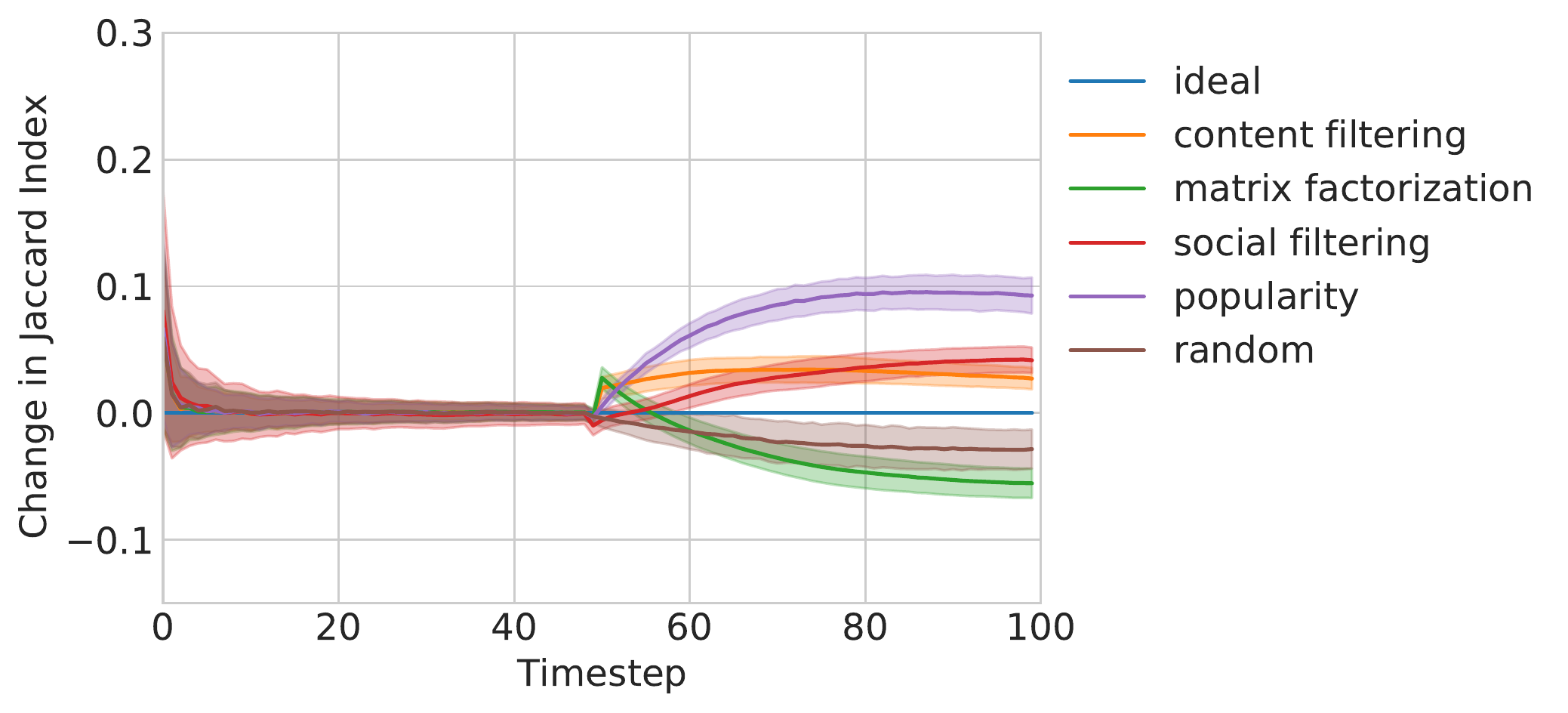}
    \caption{Our results (single training, users paired by similarity).}
    \label{fig:chaney_repeated_sim_pair}
\end{subfigure}
\begin{subfigure}{.5\textwidth}
    \includegraphics[width=\linewidth]{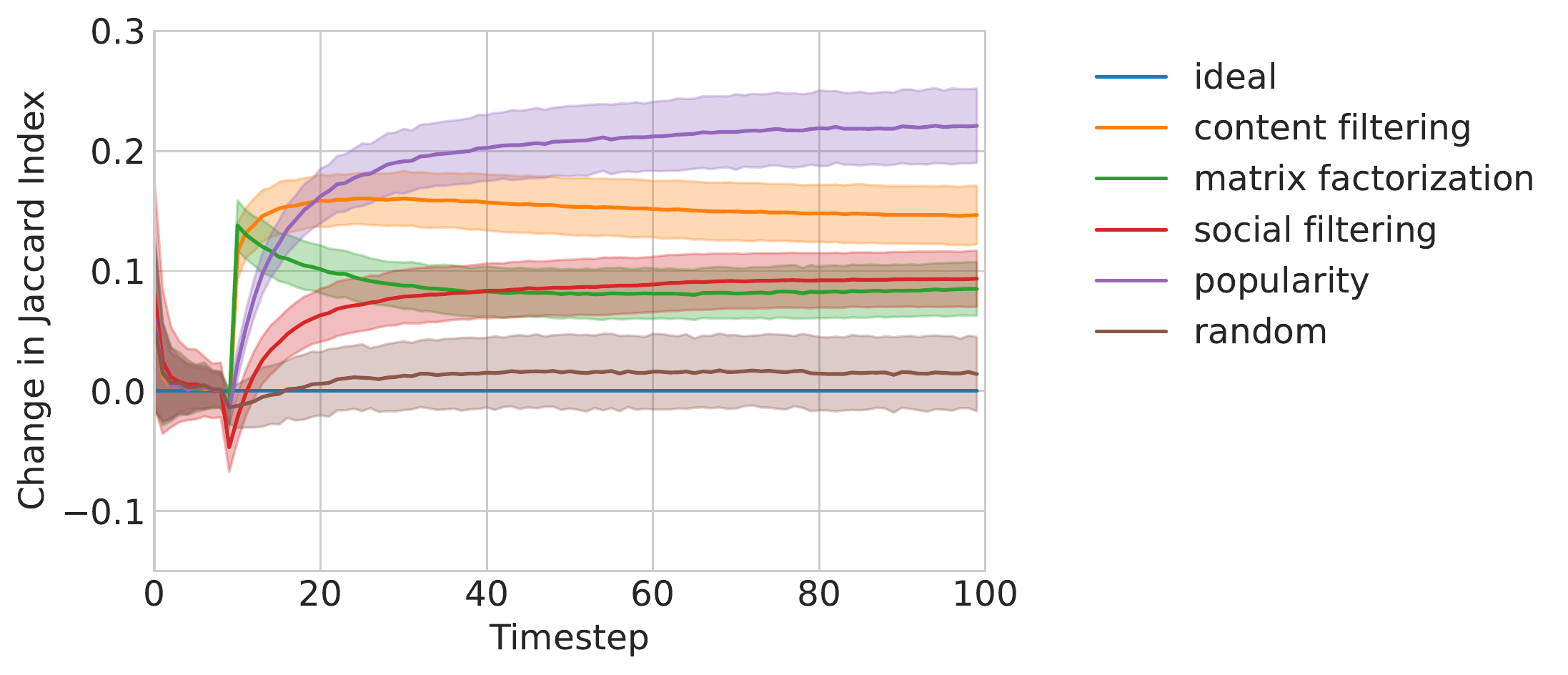}
    \caption{Our results (repeated training, users paired by similarity).}
    \label{fig:chaney_repeated_random_pair}
\end{subfigure}
\caption{Subfigures (b) and (c) show changs in Jaccard index of user behavior, users paired by cosine similarity of predicted user attributes, $\pm1$ SD across runs. We observe mild homogenization in the single training case but observe increased homogenization with repeated training. Compare to subfigure (a), a copy of the original figure from Chaney et al. \cite{chaney2018}.}
\label{fig:chaney_repeated}
\end{figure}

\subsubsection{Simulation parameters}

The simulation parameters that follow are taken from the methods of Chaney et al. \cite{chaney2018}; we describe them here for exposition. 

We run each simulation for a total of 100 steps. At each time step, we introduce 10 new items, for a total of 1000 items at the end of the simulation.

To isolate the effects of algorithm confounding, we execute the simulations in \textit{single training} and \textit{repeated training} modes. In both modes, the recommendations operate in \textit{startup mode} at first, meaning they recommend items randomly in order to gather data about user behavior. 

In single training mode, the model in each simulation is trained once after 50 \textit{startup} steps. After training, the only items recommended to users are those from the startup period (although users will see new items because of random interleaving). In repeated training mode, the models are trained at each time step after 10 startup steps. These models update their internal representations of users, item, and user-item scores at each training step. Each recommendation model provides recommendations for all items in the system at the time of training.

We differ from Chaney by running 400 trials for each recommendation model in both the single training and repeated training modes, in order to average out random noise. In each trial, the underlying user profiles, item attributes, true user-item utilities, and the user-item utilities known to the user were kept the same for all recommendation models.

\subsubsection{Metrics}

To assess the homogenization of user behavior, Chaney et al. calculate the average Jaccard index of user interactions over pairs of users. For a given pair of users, the Jaccard index is calculated as the size of the set of items both users have interacted with, divided by the size of the set of items either user has interacted with. To construct pairs of users, the authors pair each user to the user who is most similar according to the recommendation model's internal representation of user preferences. At each timestep, the average Jaccard index is measured relative to the ideal model's average Jaccard index calculated over the same pairs of users to allow for an evaluation of homogenization over and above what would occur if the users were provided with perfect recommendations.

\subsubsection{Results}
Figure \ref{fig:chaney_repeated} shows the evolution of user behavior homogenization under four different recommendation models, where homogenization is measured relative to the ideal recommendation system. We observe results that are qualitatively similar to the original study. First, our results show that the magnitude of homogenization is much greater when recommendation models are trained repeatedly. Second, in the single training case, most recommendation models exhibit an increase in homogenization directly after the training step, followed by a decrease or slowed increase in homogenization. Third, the random model homogenizes user behavior similarly to the ideal recommendation. Overall, although our results do not match Chaney et al.'s \cite{chaney2018} exactly, we emphasize that the takeaways are qualitatively similar even though the implementations (i.e., operationalizations) are slightly different. 

\subsection{Structural virality (Goel et al.)}
\subsubsection{Background.} Goel et al. use a simulation-based approach to investigate how well simple theoretical models of social contagion can capture patterns of online information diffusion observed empirically in a dataset of billions of events on Twitter \cite{goel2016structural}. They find that a relatively simple theoretical model simulated at similar scale recapitulates many of the trends found in their empirical dataset. In particular, they examine patterns of popularity--how often information cascades reach large numbers of users--and structural virality--a metric that distinguishes between viral and broadcast events. They discovered that content that is popular on Twitter is not always ``viral,'' as was often assumed. Instead, content commonly achieved popularity by being reshared by accounts with large followings. 

\subsubsection{Motivation.} We chose to replicate Goel et al.'s study for two primary reasons. First, we designed T-RECS to be general enough to accommodate algorithmic systems other than recommendation systems. Goel et al. do not explicitly attempt to simulate a recommender system, but  instead use a simulation-based approach to better understand the possible mechanisms for the diffusion of online content. By replicating their findings, we demonstrate that our library is powerful enough to provide a common framework for many different types of simulation-based research. Second, Goel et al. study outcomes that emerge from user interactions \textit{at scale}. Our replication effort simulates outcomes on networks of 1,000,000 users, indicating that researchers can use T-RECS to evaluate large-scale phenomena.

\label{sec:sec:structural-virality}
\subsubsection{Simulation framework.} Goel et al.'s models are all based on the susceptible-infected-recovered (SIR) framework, a model of contagion that is often used to model social diffusion. In SIR models, individuals are \textit{infected} (i.e., share a piece of content), and then subsequently infect their susceptible contacts independently with probability $\beta$. After being infected, individuals \textit{recover} and are no longer susceptible to infection, nor do they infect their contacts. On a given graph with average node degree $\bar{k}$, an item with infection probability $\beta$ has the ``basic reproduction number'' $r=\bar{k}\beta$.

Similar to our case study of Chaney et al. \cite{chaney2018}, we describe the parameters of our simulations; unless otherwise stated, the parameters are the same as described in Goel et al. \cite{goel2016structural}. In our replication, we generated scale-free networks of 1,000,000 users. Goel et al. use scale-free networks of size 25,000,000; we use smaller networks (albeit on the same order of magnitude) due to space and compute constraints. The degree of each graph is determined by $\alpha$, the parameter for a power law sequence. In accordance with Goel et al., we analyzed graphs for the following values of $\alpha$: 2.1, 2.3, 2.5, 2.7, and 2.9. For each value of $\alpha$, we generated 25 1M-node graphs .\footnote{Although not specified in the original paper, Goel et al. also ran multiple simulations on pre-generated graphs; we are not certain of the number of unique graphs they generated for their experiments.} Following Goel et al., in each simulation, we randomly selected a ``seed user'' to be infected with a particular item that, on a given graph, has a reproduction value $r$. We tested the following values of $r$: 0.1, 0.3, 0.5, 0.7, and 0.9. In total, we ran approximately 2,100,000 simulations across all values of $\alpha$ and $r$. Subsequently, we measured the probability of popular cascades (i.e., simulations where at least 100 users became infected) and the structural virality of these popular cascades.

\subsubsection{Results.} 
\begin{figure}[!t]
\begin{subfigure}{.5\textwidth}
    \includegraphics[width=0.8\linewidth]{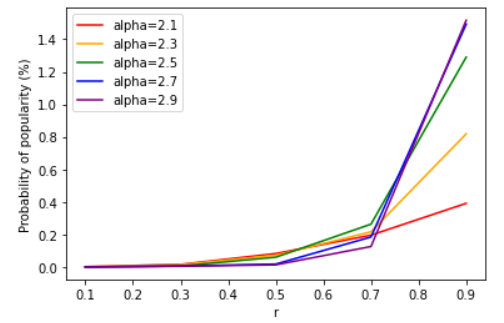}
    \caption{Chance of a cascade becoming popular (>100 users).}
    \label{fig:chaney_repeated_sim_pair}
\end{subfigure}
\begin{subfigure}{.5\textwidth}
    \includegraphics[width=0.8\linewidth]{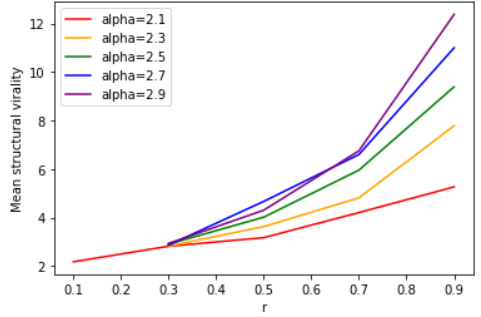}
    \caption{Mean structural virality of popular cascades.}
    \label{fig:chaney_repeated_random_pair}
\end{subfigure}
\caption{Probability of popular cascades and mean structural virality for SIR simulations on random scale-free networks. Compare to Figure 7(A) and 7(B) of Goel et al. \cite{goel2016structural}}
\label{fig:goel}
\end{figure}

The results of our simulations are shown in Figure \ref{fig:goel}. Our findings are qualitatively similar to Goel et al.'s, even though our simulations were performed on 1M node graphs, while theirs were performed on 25M node graphs. For higher values of $r$, we observe higher probability that a given cascade reaches at least 100 users. We also find that at lower values of $r$, content is less likely to become popular on graphs generated with lower values of $\alpha$, and at higher levels of $r$, the opposite is true. In line with Goel et al.'s findings, we also observed that the mean structural virality was generally highest for high values of $\alpha$ and increases with $r$.

Finally, Goel et al. observed that simulations run with parameters $r\approx 0.5$ and $\alpha\approx 2.3$ best matched the patterns in the empirical Twitter dataset. For this parameter setting, they found that the probability of a content becoming popular was about one in a thousand, and the mean structural virality of popular cascades was about 3.7. This matches the empirical results from our simulations as well. Goel et al. also find that the correlation between the size of popular cascades and structural virality is approximately 0.1, which falls within range of 0 to 0.2 which was observed in the Twitter data. Similarly, our simulations with the parameter setting $r\approx 0.5$ and $\alpha\approx 2.3$ yielded a correlation of $\approx 0.086$ between size and structural virality.

\subsection{Challenges in replicating simulation-based research}
The prior two case studies illuminated some challenges in replicating simulation-based research. In theory, the methodology section of a technical report should be sufficient for another researcher to be able to fully replicate the design. In practice, few research designs are simple enough to be completely described within a relatively short methodology section. To reproduce the results of Chaney et al. \cite{chaney2018} and Goel et al. \cite{goel2016structural}, we needed to consult not only the methodology section, but also methodological appendices, footnotes, and details that appeared only in the graphs of results. When the researcher's primary goal is explaining the theoretical model to the reader, focusing only on the most critical details in the methodology section is understandable; however, this approach has a side effect of rendering the study more difficult to replicate. To successfully produce our conceptual replications, our team frequently had meetings to parse the exact meaning of single sentences or even single phrases within the text to attempt to correctly grasp their intended meaning. Even then, there were multiple plausible operationalizations of the theoretical models described. For example, in the single training case of Chaney et al.'s study, we were not certain whether new items were permanently inserted in each user's recommendation list after random interleaving, or if the new items interleaved at timestep $t$ were never interleaved again at any later timestep. Such details are incidental to the high-level results of the paper; however, we found that such details were critical for correctly reproducing the results. The accumulation of ambiguities in seemingly trivial implementation details meant that our research team spent considerable time and effort conducting analyses to rule out many possible specifications before arriving at the specification that produced results most similar to the original research. Using T-RECS allows researchers to use the language of our framework to describe their methodology in terms that T-RECS users will be familiar with. Adoption of T-RECS also encourages researchers to upload and share code from their experiments -- which would completely eliminate the aforementioned ambiguities -- since code that uses a common framework - as opposed to an ad-hoc solution - is likely to be more useful and legible to other researchers. 

Second, the large scale of the simulations in Goel et al. \cite{goel2016structural} presented several computational challenges. Without a clear description the resources required to perform the original study, our team needed to experiment with several different hardware configurations and implement software optimizations to replicate their findings. Future simulation studies can better facilitate replication by providing a description of the hardware and memory requirements for conducting their analysis. In doing so, other researchers would be better able to determine whether whether a replication or novel extension of the work would even be feasible. Similarly, sharing descriptions and code regarding various optimizations techniques (for example, the use of sparse matrices) would likewise reduce the time and effort necessary for other research teams to reproduce -- and extend -- original simulation findings. Again, we see T-RECS's role in encouraging the sharing of reproducible simulation code as naturally mitigating this issue, as researchers can simply run the simulation code on their own machines to measure performance, without having to rebuild the ad-hoc system. Furthermore, contributors to T-RECS will continue to improve the library, including adding speed and memory optimizations, allowing the entire research community to benefit and removing the need for isolated research teams to reinvent the wheel. For example, support sparse matrices is currently included by default in T-RECS.

Third, rarely do papers describe their procedure for testing the correctness of their implementations; instead, correctness of implementation is taken as a given. We say this not to imply that simulations from prior research were incorrect in any way -- rather, we aim to highlight the difficulty of ensuring there are no implementation errors in a simulation of a large, complex system, particularly in cases where the simulation environment is being built ad hoc. Like many machine learning systems, these simulations can fail \textit{silently}, meaning that they do not raise obvious bugs or errors during runtime. Instead, the researcher observes outcomes that are obviously incorrect, and then must work backward through the entire simulation implementation to understand what has gone awry. With this issue in mind, we designed T-RECS with thorough test coverage, believing that well-tested code is essential to robust, reproducible simulation-based research.

Finally, future simulation-based research should include confidence intervals on all results. For example, where figures are shown with results averaged over multiple simulation trials, it would be ideal to provide information about the variance of the results across trials. This would aid future researchers in understanding whether their observed results differ from those reported in the original paper because of differing implementations/assumptions or statistical noise. In our case, this information would have provided an additional amount of confidence that we had properly replicated previous findings, despite minor differences in the appearance of our generated figures.

\subsection{Content creators and polarization}
For our final case study, we present novel work examining a different question: how does the presence of adaptive content creators affect the distribution of \textit{dynamically generated items}, in comparison to when items are served from a fixed catalog? To answer this question, we set up our environment with nearly identical assumptions to those of Chaney et al. The key difference is that the new items introduced at each timestep, rather than being sampled randomly from a fixed distribution, are generated by a pool of content creators, each with their own time-varying item-generating distributions. At each timestep, these content creators adapt their item-generating distributions to user feedback at each iteration. We investigate how this process results in changes to the items generated over time, when each creator starts out by sampling items from a fairly uniform distribution across item attributes.

Importantly, we reuse much of the code and many of the assumptions from the case study of Chaney et al. In the original Chaney simulations, new items are introduced into the system at each timestep, but they are sampled from a fixed distribution. When new items are instead produced by content creators who respond to user feedback, we may observe that creators themselves are ``polarized'' in that they begin the simulation by generating a diverse range of items, but then adapt to feedback by generating items from narrower and narrower subsets of the space of possible items.

Our reuse of code affords two main benefits  that are indicative of the advantages of using T-RECS in general. First, it mitigates the possibility that differences observed are due to differences in implementation of the simulation environment, rather than the presence of content creators. Second, from a practical standpoint, it saves us a great deal of time since we no longer have to re-implement the assumptions from the original Chaney et al. sutdy from scratch.

\subsubsection{Measuring creator homogenization}
In the original study by Chaney et al., homogenization is measured through an average Jaccard index of sets of user-item interaction histories \cite{chaney2018}. In our study, we shift our focus to studying the homogenization of creators, rather than the homogenization of users. We use the average entropy of each creator's item-generating distribution as a rough proxy for homogenization, henceforth referred to as average creator entropy (ACE). Intuitively, if the entropy of all creator item distributions is high, then each creator is likely to produce a diverse set of items, and if the entropy of the item-generating distributions is low, then each creator is generating a group of items that are more uniform in their attributes. Note that this measure captures \textit{within-creator} homogenization; that is, it measures whether each creator is sampling from a narrow or broad distribution of items, not whether creators are increasingly similar to each other. Therefore, this definition of homogenization does not imply that there cannot be diversity across creators as a whole; instead, it describes individual creator behavior.

\subsubsection{Adaptive content creators.} The specific item-generating distribution from which content creators sample is a Dirichlet distribution. Each creator's attributes $\gamma_i$ is initially sampled in the following manner: $\gamma_i \sim \text{Dirichlet}(\mathbf{10})$. For a given creator $i$, the item-generating distribution is $\alpha_i \sim \text{Dirichlet}(\gamma_i \cdot 0.1)$. This ensures that item attributes are sparse and are highly sensitive to shifts in the creator's attributes profile, which is initially spread somewhat evenly across most attributes. Recall that this is very similar to how the static item-generating distributions are sampled in Chaney et al. \cite{chaney2018}.

Content creators respond to feedback by shifting their profiles $\gamma$ towards items they produced with which users interacted during the most recent timestep. (Note that both creator profiles and item attribute vectors sum to one, since both are sampled from Dirichlet distributions.) Notably, we do not claim that this is the best or only way to model adaptive behavior; instead, we posit it as a possible way to approximate how content creators respond to the incentive to maximize user interactions with their content.

\subsubsection{Results}
\begin{figure}[!t]
\includegraphics[width=\linewidth]{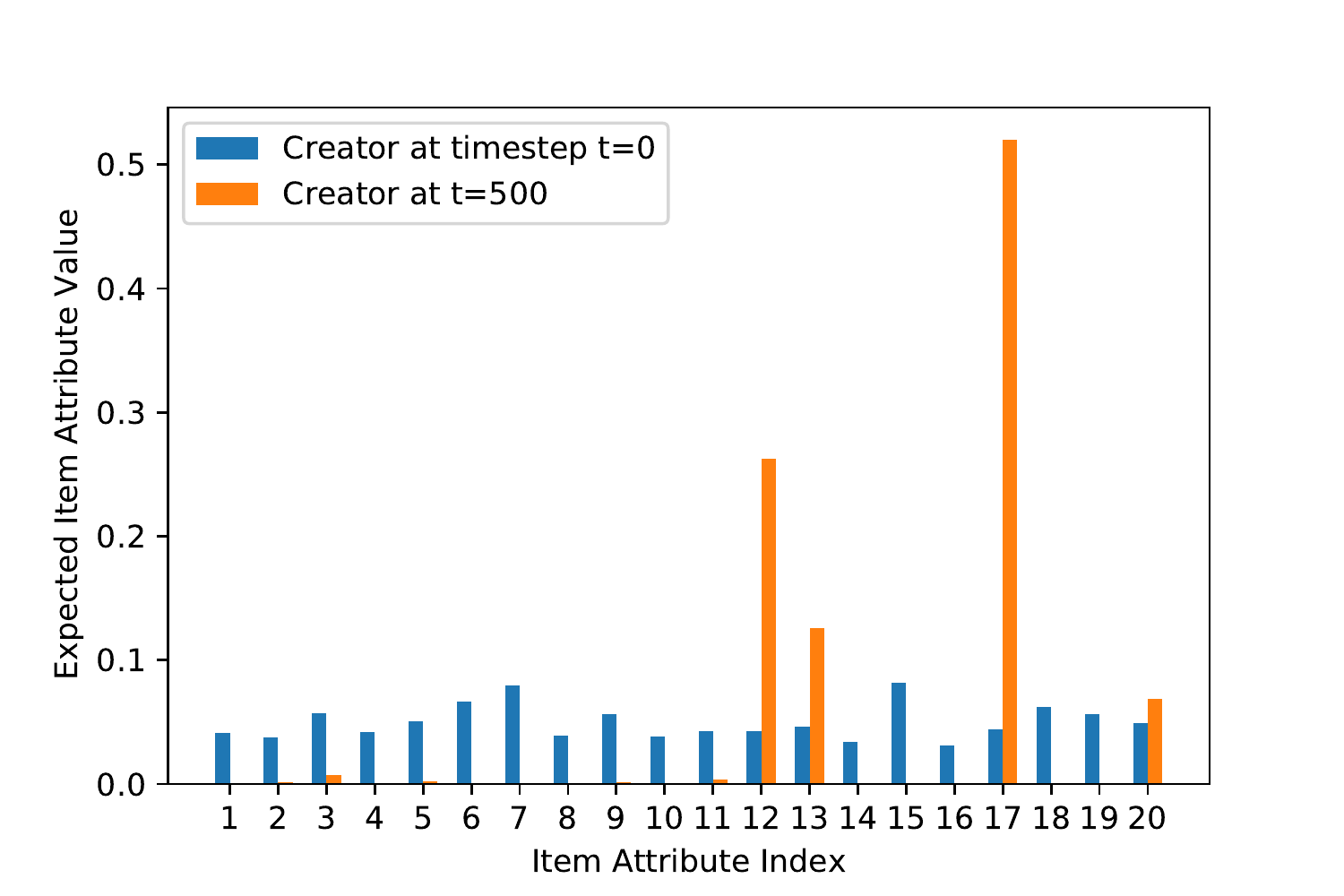}
\caption{To illustrate the phenomenon of creator homogenization, we show the change in a randomly chosen creator's item-generating distribution at the start and end of the simulation. At $t=0$, the expected item attribute values are all approximately the same. Since items are sparse, this indicates that the creator is equally likely to create an item with any of the possible attributes. At $t=500$, however, the creator is likely to generate only items that have a small set of attributes.}
\label{fig:creator_avg_item}
\end{figure}

In Figure \ref{fig:creator_avg_item}, we provide a visual representation of the trajectory of a single creator's item-generating distribution from the beginning to the end of the simulation. The recommendation system used was the ``ideal'' recommender from the Chaney et al. study. We observe that at the beginning of the simulation, the creator is about equally likely to create an item with any of the possible attributes. However, at the end of the simulation, the creator has ``narrowed'' to generating items that have a much smaller set of possible attributes; the creator essentially never generates items that have a high value for the other attributes.

\begin{figure}[!t]
\includegraphics[width=\linewidth]{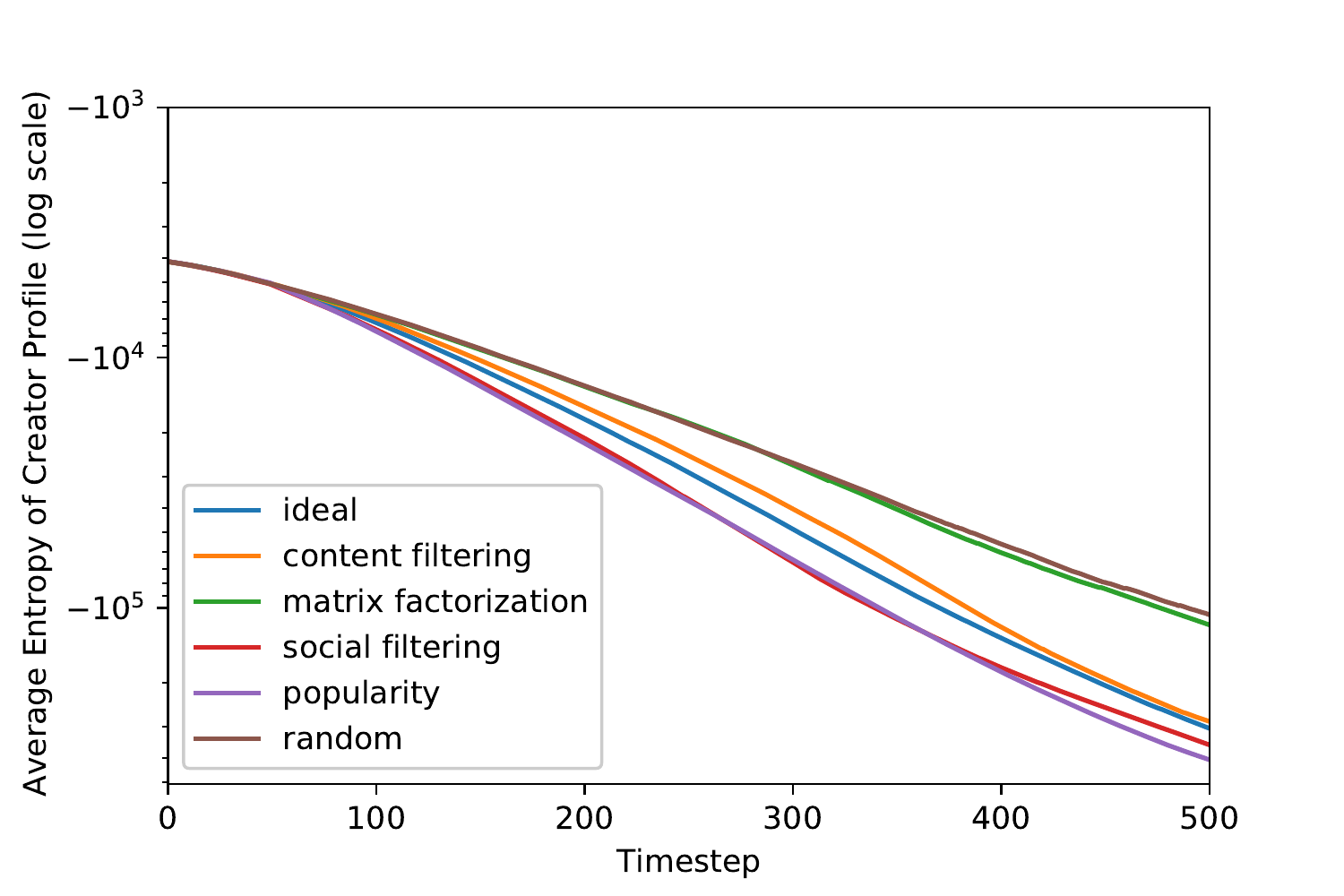}
\caption{Average entropy of the creators' item-generating distributions across. For all systems, including the random recommendation scheme, entropy decreases over time, indicating that each individual creator's item-generating distribution is being homogenized over the duration of the simulation. Note that since we use a logarithmic scale on the y-axis, the rate of decrease accelerates over time.}
\label{fig:creator_entropy}
\end{figure}

To illustrate this phenomenon across all recommender systems, we plot the ACE at each timestep in Figure \ref{fig:creator_entropy}. In all recommender systems, ACE decreases throughout the simulation at an increasing rate. We also observe a greater degree of creator polarization in the social filtering and popularity recommender systems, suggesting that algorithm selection makes a difference in the rate or degree to which creators are homogenized.

Finally, we performed a preliminary exploration of how creator homogenization impacts user homogenization, using a different measurement of user homogenization than the Chaney et al. use in their study. Our proxy for user homogenization is averaged over pairs of users, where the metric averaged is the distance between the mean items interacted with by the two users. Formally, our metric is: $\frac{1}{n} \sum_{k, j} d(\bar{i}_k, \bar{i}_j)$, where $n$ is the number of pairs of users, $\bar{i}_u$ is the average item profile across user $u$'s interaction history, and $d$ is the Euclidean distance function. We refer to this measure as Average Pairwise Distance between Mean Consumed Items (APDMCI). We find that for all non-ideal recommender systems, APDMCI increases beyond what is necessary to achieve optimal utility for users. This suggests that as individual creators become homogenized, user behavior is also homogenized. 

Note that the Jaccard-based measure of user homogenization is a reasonable measure to use when the item-generating distribution is fixed, but is not able to capture an important dimension of homogenization when creators adapt to user feedback. Consider the example where User A and User B are loyal consumers to Creator C and Creator D, respectively. Initially, Creators C and D generate items that are extremely different from one another, but over time both creators ``drift'' towards each other, creating items that differ only moderately. In this scenario, a Jaccard-index based measure would report no change in homogenization throughout the experiment, although users A and B are now consuming items that are not as different from each other as they used to be.
 
\subsubsection{Discussion}
We observe that when content creators adapt to user behavior, within-creator homogenization may occur, as measured by increasing ACE. This means that over time, creators respond to the engagement incentive by narrowing the set of items they produce. 

First, we stress that we do not make a normative claim that within-creator homogenization is a negative outcome. In certain domains, our definition of homogenization can be interpreted as creators learning users' interests and responding to user interests. We suggest that recommender system designers consider designing domain-specific interventions to mitigate the possible consequences of polarization. 

Second, we also note that we observed a high degree of variance between runs. We average our results over hundreds of simulations to reduce variance in our calculation of the mean outcome. Sources of intrinsic variance include the sampling of new user and creator profiles for each simulation and the noise that is internal to each recommender system (for example, the process of randomly interleaving new items into recommendation sets). The ability to run this many repeated experiments is a strength of the simulation approach; it may be comparatively expensive or cumbersome to repeatedly sample new users and retrain recommender systems in real-world settings.  

Finally, our study shows the need for clarity in how the research community defines concepts like homogenization and polarization that can apply to a wide range of settings. By using a common framework like T-RECS, we can hold all components of an experiment constant and compare how different metrics might capture fundamentally different phenomena. For example, we discovered during our experiments that the measure of user homogenization used by Chaney et al. did not translate perfectly to the adaptive content creator setting, and thus formulated our own measure \cite{chaney2018}. 

These results are intended as an initial exploration into the role that content creators play in recommender systems. Different models of content creators and how they adapt to user feedback may lead to different results. As a result, we invite further research into the interplay between content creators, users, and recommendation system algorithms.

\section{Conclusion} \label{sec:conclusion}
Recommender systems comprise an increasingly central role in many of our collective and individual digital experiences, from reading the news, watching movies, and connecting with friends. In part because recommender systems deployed in production are proprietary and also vary significantly by use and application, researchers in the social sciences and computer science have turned to simulation-based approaches. However, most of these efforts have required researchers to build ad-hoc systems, leading to significant and redundant engineering burdens across studies. Moreover, as recommender systems become more complex, the time required to build even a prototype ad hoc simulation will only increase, as will the likelihood of bugs and errors. 

Informed by both an examination of the current trends in simulation-based recommender systems research, T-RECS offers unified, flexible framework for simulating the dynamic interplay between users, items, and algorithms in recommender systems and other sociotechnical systems. Adopting T-RECS offers benefits for individual practitioners, including the guarantee of well-tested code and significant reductions in engineering burden, as well as for the research community as a whole, including the promotion of reproducibility, collaboration between researchers, and a common language for reasoning about complex phenomena in recommender systems.

\section*{Acknowledgments} \label{sec:acknowledgments}
This work is partly supported by the 2021 Twitch Research Fellowship and the 2021-22 Hamid Biglari *87 Behavioral Science Fellowship.
We are also grateful for support from the National Science Foundation under Award IIS-1763642. Additionally, we gratefully acknowledge financial support from the Schmidt DataX Fund at Princeton University made possible through a major gift from the Schmidt Futures Foundation.


\bibliographystyle{ACM-Reference-Format}
\bibliography{bibliography}

\appendix
\section{Appendix} \label{sec:appendix}

\begin{table*}[t]
\begin{tabular}{p{0.15\textwidth}p{0.2\textwidth}p{0.65\textwidth}}
\hline
\multicolumn{3}{c}{\textbf{Simulation to study sociotechnical systems}}                                                                                                                                                                               \\
\textbf{Work}                                                         & \textbf{Concept}                        & \textbf{Metric(s)}                                                                                                                  \\ \hline
\multirow{4}{*}{Aridor et al.~\cite{aridor2020deconstructing}}        & \textcolor{red}{Filter bubble}          & Average consumption distance at time $t$ and $t-1$                                                                                  \\
                                                                      & User welfare                            & Average realized utility                                                                                                            \\
                                                                      & Item diversity                          & Average normalized pairwise distance between consumed items as seen in \cite{ziegler2005improving}                                  \\
                                                                      & User homogenization                     & Average pairwise Jaccard index (as seen in \cite{chaney2018})                                                                       \\ \hline
\multirow{2}{*}{Chaney et al.~\cite{chaney2018}}                      & User homogenization                     & Jaccard index on the sets of seen items by pairs of users                                                                           \\
                                                                      & Differential item consumption           & Gini index on the distribution of consumed items                                                                                    \\ \hline
\multirow{2}{*}{Ciampaglia et   al.~\cite{ciampaglia2018algorithmic}} & User welfare                            & Average quality of items                                                                                                            \\
                                                                      & Faithfulness                            & Kendall rank correlation between popularity and quality of items                                                                    \\ \hline
Garimella et al.~\cite{garimella2017balancing}                        & \textcolor{red}{Filter bubble}           & Expected number of users exposed to both or neither of two items (representing viewpoints)                                          \\ \hline
Geschke et al.~\cite{geschke2019triple}                               & \textcolor{blue}{Echo chamber}          & \begin{tabular}[c]{@{}l@{}}Mean distances between users and items/item sharers/friends\\ Visual analysis of clustering\end{tabular} \\ \hline
Goel et al.~\cite{goel2016structural}                                 & Structural virality                     & Wiener index of diffusion tree \cite{mohar1988compute}                                                                              \\ \hline
\multirow{2}{*}{Jiang et al.~\cite{jiang2019degenerate}}              & \textcolor{blue}{Echo chamber}          & Distance between initial user interests and final user interests (weak or strong degeneracy)                                        \\
                                                                      & \textcolor{red}{Filter bubble}          & Speed of degeneracy                                                                                                                 \\ \hline
Lee et al. ~\cite{lee2019homophily}                                   & Social perception bias                  & Error of individuals in estimating true prevalence of binary attribute                                                              \\ \hline
Lim et al.~\cite{lim2014opinion}                                      & \textcolor{darkgreen}{Polarization}         & Number of clusters, Herfindahl-Hirschman Index (HHI), proportion of agents in majority/minority clusters, speed of convergence      \\ \hline
Nasrinpour et al.~\cite{nasrinpour2016agent}                                      & Virality         &  Number of interactions (reads, reposts)      \\ \hline
\multirow{2}{*}{Perra and Rocha \cite{perra2019modelling}}            & \textcolor{darkgreen}{Polarization}         & Prevalence of opinions over time                                                                                                    \\
                                                                      & \textcolor{blue}{Echo chamber}          & Distribution of neighbors holding the majority or minority opinion at end of simulation, relative to start of simulation            \\ \hline
\multirow{2}{*}{Sun et al.~\cite{sun2019debiasing}}                   & Accuracy                                & Root Mean Squared Error                                                                                                             \\
                                                                      & Popularity bias                         & Gini coefficient                                                                                                                    \\ \hline
\multirow{2}{*}{Tambuscio et al.~\cite{tambuscio2018network}}         & Spread of misinformation                & Fraction of infected users at equilibrium, state transition rates                                                                   \\
                                                                      & \textcolor{blue}{Echo chamber}          & Custom generative model for segregated network                                                                                      \\ \hline
\multirow{3}{*}{T\"ornberg \cite{tornberg2018echo}}                   & Virality                                & Probability that majority of nodes are "infected"                                                                                   \\
                                                                      & \textcolor{darkgreen}{Network polarization} & Increased ties within cluster, decreased ties from cluster to outside                                                               \\
                                                                      & \textcolor{darkgreen}{Opinion polarization} & Probability that neighboring nodes have similar activation thresholds                                                               \\ \hline
Yao and Huang \cite{yao2017beyond}                                    & Group unfairness                        & Four different unfairness metrics                                                                                                  
\end{tabular}
\caption{Simulation literature survey. Note that many different metrics are used to assess the dimensions of various concepts. Concepts that are similar in nature are given the same color.}
\label{tab:concept_survey}
\end{table*}

\subsection{Replication} \label{sec:appendix_replication}

\subsubsection{Algorithmic confounding (Chaney et al.)}
In the following sections, we provide more details about our methods reproducing Chaney et al.'s work \cite{chaney2018}. If we have omitted any details, please defer to the assumptions laid out in the original paper.

\textbf{Recommender system models.}
Following Chaney et al., we model six types of recommender systems. Each recommender system maintains an internal representation of predicted user preferences and predicted item attributes. Recommendations to each user are based on the predicted user scores, which are themselves a function of the predicted user preferences and predicted item attributes (typically a dot product).

The first model is content-based filtering, which recommends content similar to what users have liked in the past. As Chaney et al. indicate, the predicted user preferences in this model are updated at each step by solving for the least-squares approximation of user attributes using the \texttt{scipy.optimize.nnls} function \cite{2020SciPy-NMeth}. Item attributes are equal to the true item attributes.

Second, we use popularity-based filtering, which serves the most popular items in the systems. In this model, the predicted item attributes is equal to a single number for each item: the total number of interactions that item has received. Predicted user preferences are identical and constant for all users, so that the predicted score for user $u$ and item $i$ is simply equal to the number of interactions that item $i$ has received.

Third, we implement a matrix factorization collaborative filtering model in which a common latent representation of users and items is used to recommend items based on user interaction data. The predicted user preferences and item attributes are generated using the alternating least squares approach \cite{hu2008collaborative} implemented in the LensKit library \cite{ekstrand2020lenskit}. Note that this matrix factorization model is not identical to the model used by Chaney et al.

Fourth, we implement social-based filtering, which recommends items based on the preferences of users in their social network. In this model, predicted user preferences are represented with an adjacency matrix that includes the connections between users. Following Chaney et al., we generate this adjacency matrix from the covariance matrix of true user preferences.

Lastly, we provide two baseline models: a random recommender system and an ideal recommender system. The former serves random items to users. The latter presents items based on the users' true utility.

\textbf{Synthetic data.}
Next, we generate the user and item data. The actual user preferences $U$ are represented as a $|U| \times |A|$ matrix, where $|U|=100$ is the number of users and $|A|=20$ is the number of \textit{attributes} describing each user. The $i$-th row of $U$ contains the attributes describing user $u_i$. The values of the rows were drawn from a Dirichlet distribution with parameters as specified by Chaney et al.

Each model requires a distinct representation of the predicted user preferences $\hat{U}$. In content-based filtering, predicted user preferences are a $|{U}| \times |A|$ matrix with the same properties as the actual user preference matrix $U$. In the popularity-based recommender system, $\hat{U}$ is a $|U| \times 1$ matrix with all elements equal to 1, as the predictions of the system are the same for all users (recall that predictions are calculated with the dot product of the user item matrices). In our collaborative filtering model, we use matrix factorization; therefore, predicted user profiles are a $|U| \times k$ matrix, with $k$ being the number of features in the latent representation. In our social-based filtering model, we represent the users' social networks in $\hat{U}$; therefore, $\hat{U}$ is a $|U| \times |U|$ matrix.

Items in the system are represented by a matrix $I$ of size $|I| \times |A|$, where $|I|=1000$ is the total number of items and $|A|=20$ is the number of attributes that describe each item, analogously to matrix $U$ for users. The values of the rows were drawn from a Dirichlet distribution with parameters as specified by Chaney et al.

We additionally calculate users' \textit{utility} as defined by Chaney et al. We specifically distinguish between true utility and \textit{known} utility. The latter is the fraction of true utility known to users and is equivalent to the actual user scores in our framework --- that is, the scores that the recommender systems predict. 

True and known utilities are represented with two distinct matrices of size $|U| \times |I|$. For a given user $u$ and a given item $i$, the true utility obtained by $u$ interacting with $i$ is sampled from a Beta distribution whose mean is parameterized as the dot product of the $u$'s true preferences and $i$'s true attributes. From the vantage point of $u$, the known utility of interacting with $i$ is a percentage drawn from a beta distribution with parameters as specified by Chaney et al.

\textbf{Performance} The total time to run the entire set of experiments was just under 3 hours on a laptop with 32 GB of RAM and a 6-core 2.6 GHz processor.

\subsubsection{Structural virality (Goel et al.)}

\textbf{Performance.} Several optimizations were required for our replication effort. First, we had to ensure that the large graphs of millions of users fit into memory during simulation. T-RECS supports \texttt{scipy} sparse matrices \cite{2020SciPy-NMeth}, which allowed our simulations of hundreds of thousands of users to run using just a few gigabytes of RAM. Second, we made thorough use of multiprocessing-based parallelization to generate tens of thousands of simulation runs within days; this was essential because the probability of generating a popular cascade on any given run was extremely low (approximately 0.1\%). Examples of using multiprocessing to achieve these speedups are available online at \href{https://github.com/sunnymatt/t-recs-experiments}{https://github.com/sunnymatt/t-recs-experiments}.

\subsubsection{Content creators}

\textbf{Performance.} We perform 200 separate 500-timestep trials for each of our six recommendation algorithms in approximately six hours. While our simulations were performed on our university's academic computing cluster, the simulations utilized just 4.20 GB of memory, suggesting that the experiments could also be performed on regular desktop or laptop hardware.


\subsection{Internal representation for users and items} \label{sec:sec:internal}
\textbf{Challenge: developing a scalable and flexible representation to encode information about users and items.}

We considered two different engineering approaches to represent users and items internally:

\begin{enumerate}
    \item Dedicate a data structure to each user and item.
    \item Dedicate one data structure to all users and another data structure to all items.
\end{enumerate}

While the first approach may be the most natural choice for an object-oriented design and well-suited to emphasizing the unique characteristics of each user and item, the latter has several advantages. First, it scales well with the addition of  more users or items. Second, it is flexible because the system does not need to make assumptions on the number of data structures dedicated to users and items, as it will be constant to one. Third, it allows to process user actions on items in parallel.

As seen in previous work \cite{chaney2018}, T-RECS stores information about all the users in the system in one matrix; similarly, all the items are represented with a single matrix. For both users and items, two representations are generated---a ground truth representation and a predicted representation. Therefore, we use a matrix to represent predicted user preferences, and a separate matrix for the actual user preferences.

\subsubsection{Actual user preferences.} Bidimensional matrix $U$ with the first dimension equal to the number of users in the recommender system ($|U|$). The second dimension varies with the model. Typically, the $i$-th row represents the preferences of user $U_i$ ($i \in \{0,\dots, |U|-1\}$). Actual user preferences are unknown to the recommender system and only known to the users.
\subsubsection{Predicted user preferences.} Bidimensional matrix $\hat{U}$ with the same properties as $U$. This matrix is calculated by the model and represents inferred user preferences.
\subsubsection{Item attributes.} Bidimensional matrix $I$ with the second dimension equal to the number of items in the recommender system ($|I|$). The first dimension varies with the model. Typically, the $j$-th column represents the attributes of item $I_j$ ($j \in \{0,\dots, |I|-1\}$). For simplicity, we assume that the model knows the real item attributes; therefore, predicted item attributes are equal to the actual item attributes.

\subsubsection{Actual user scores.} Matrix $S$ of dimensions $|U|\times|I|$. The element at $S_{ij}$ represents the ground truth score for user $U_i$ on item $I_j$. These scores are typically used to define the user behavior when giving feedback to items that are presented to them. That is, users will interact with items with the highest predicted actual user score. T-RECS provides two built-in methods to calculate actual user scores: the inner product between actual user preferences and item attributes, or the cosine similarity between the two matrices. Actual user scores are unknown to the recommender system and are only known to the users.
\subsubsection{Predicted user scores.} Matrix $\hat{S}$ of dimensions $|U|\times|I|$. Predicted user scores have the same dimensions as $S$, but are is calculated by the model using $\hat{U}$ instead of $U$. Predicted user scores are used by the model to recommend item to users.

\subsubsection{Metrics}
\textbf{Challenge: instrumenting the simulator to gather information over time about.}

A fitting implementation choice for metrics was the observer design pattern, a standard pattern in object-oriented programming that is used when one object (the \textit{observer}) needs to be ``notified'' automatically when a second object (the \textit{observable}) changes. In the case of T-RECS, the model is the observer and the metrics are the observable objects updated at each step. Specifically, we define the observer paradigm with a single observer and multiple observables, as opposed to the traditional pattern with multiple observers and a single observer. Therefore, models can \textit{register} (i.e., monitor) and unregister (i.e.,, stop monitoring) multiple metrics. Using the observer design pattern also provided flexibility to easily develop and integrate new metrics.

Based on a review of the simulation literature about algorithmic systems (Table \ref{tab:survey}), we identified two different kinds of metrics: \textit{measurements} and internal \textit{system states}. They are both implemented following the observer pattern.

Measurements are designed to calculate quantities of interest about the system as a whole. For example, diversity of content can be calculated by comparing the distances between the user preferences and the attributes of the items recommended \cite{geschke2019triple}.

System states expose relevant information about the simulation and the system. Unlike measurements, system states are presented as is and do not require additional computation. For instance, researchers interested in monitoring the evolution of the predicted user preferences can register matrix $\hat{U}$ as a system state; at each iteration, the state of $\hat{U}$ will be stored internally. This also provides a mechanism useful to debug model behavior and retrieve quantities for later inspection and manipulation.

\subsubsection{Content creators}
By default, content creators are modeled by a matrix $\mathcal{C}$ of dimension $\mid C \mid\times\mid A\mid$, where $C$ is the number of content creators and $A$ is the number of attributes that parameterize each creator's item-generating distribution. In the default model, each content creator $c$ is modeled as a vector of length $A$, where each entry in the vector denotes $p_a$, the Bernoulli probability that an item created by $c$ will have a value of 1 for attribute $i$. (Note that this implies that item attributes are sampled independently.) Researchers can also pass in an optional probability $p_c$ that represents the probability that each creator generates an item at any timestep; this quantity can be used to determine the general ``productivity'' level of the content creator pool.

We aimed to maximize simplicity and modularity with our implementation of content creators. A simulation with 100 content creators that generate items with 20 attributes can be instantiated as follows:

\begin{verbatim}
c = Creators(size=(100, 20))
recsys = trecs.models.ContentFiltering(creators=c)
recsys.run(timesteps=100)
\end{verbatim}

This ``drop-in'' API allows researchers to easily measure the impact of adding content creators to their simulations. 

This simple default setup is not intended to be highly realistic; for example, it does not impose any restrictions on the sparsity of item attributes. Instead, the default implementation is meant to illustrate the general mechanics of representing content creators through the matrix $\mathcal{C}$, which contains the parameters for each content creator's item-generating distribution. Because the research community has not yet reached consensus on the correct theoretical model for content creators, we expect researchers to test theoretical variations of content creators by implementing their own content creator subclasses or by modifying the existing content creator subclass to better suit their research questions.

Simulating content creators generally induces a greater computational cost per timestep than a fixed item catalog, since the new items are sampled at each timestep. As the item set grows over time, long simulations may result in high memory and compute requirements. The choice of item-generating distribution also affects simulation performance; for example, we found in practice that modeling each content creator's item-generating distribution as a multivariate normal slowed simulations by an order of magnitude.

Finally, the basic simulation dynamics described in Section \ref{sec:sec:dynamics} can also be modified with the following options:

\subsection{Dynamic user preferences} Modeling how users themselves change over time is key to capturing phenomena of interest that occur in some part at the level of individual users, such as political polarization or ideological radicalization. For example, Geschke et al. \cite{geschke2019triple} simulation-based studies model how agents attitudes' shift over time in environments with or without recommendation algorithms. Jiang et al. \cite{jiang2019degenerate} provide a theoretical model for how recommender systems might alter user preferences over time. In their model, the utility a user receives from interacting with an item at timestep $t$ might depend on the user's history of interactions at all previous timesteps $t-1,t-2,\cdots$. Empirical research also seems to suggest that user's interest might change over time, such as when users gravitate to progressively more extremist content on YouTube \cite{ribeiro2020auditing}. The capacity to model how individual users are affected by their interactions with sociotechnical systems is necessary to capture these types of effects on users.

In T-RECS, choosing to model dynamic user preferences in a given simulation translates to an added step after users choose which item to interact with from the recommendation set. For a given user, the user's attributes ``drift'' towards the chosen item's attributes; concretely, we implement spherical linear interpolation \cite{shoemake1985animating}, such that the user's profile vector is rotated in the direction of the item vector. Users of the T-RECS library may also choose to implement their own custom drift functions.

\subsection{Start-up mode} Many recommendation models are unable to make recommendations to new users (i.e., \textit{cold-start}); therefore, we provide a mechanism for the system to gather information about users’ preferences during a \textit{start-up phase}. We assume that users have preferences that can be expressed for all items; therefore, we present items randomly during start-up to maximally sample the range of users’ preferences and minimize the risk that start-up preference elicitation strategy could bias the model. 

As with other system components, researchers can define their own start-up strategies if they are interested in the effect of start-up behavior on subsequent system state. For example, a reasonable assumption is that users cannot express preferences for unfamiliar items, so practitioners interested in collecting start-up preferences as efficiently as possible may benefit from showing users popular items \cite{rashid2002getting}. However, popular items are the least informative about individual users’ preferences since they tend to be preferred by all users.

In start-up mode, the system skips step 1 of the simulation dynamics. At the end of the start-up phase, the model is trained on the collected interactions -- that is, the predicted user scores are calculated based on the items users have interacted with. Note that the T-RECS API does not support adding and removing users during a simulation, so the cold start problem only manifests itself when training a new model from scratch.

\end{document}